\documentclass[showpacs,amsmath,amssymb, prb,twocolumn]{revtex4}
\usepackage{graphicx}
\usepackage{dcolumn}
\usepackage{bm}

\begin{document}

\title{Valley to charge current conversion in graphene grain boundaries}

\author{Francesco Romeo}
\affiliation{ Dipartimento di Fisica ''E. R. Caianiello'',
	Universit\`a degli
	Studi di Salerno, Via Giovanni Paolo II, I-84084 Fisciano (Sa), Italy}

\begin{abstract}
The conduction properties of a grain boundary junction with Fermi velocity mismatch are analyzed. We provide a generalization of the Dirac Hamiltonian model taking into account the Fermi velocity gradient at the interface. General boundary conditions for the scattering problem are derived within the framework of the matching matrix method. We show that the scattering properties of the interface, as predicted by the theory, strongly depend on the boundary conditions used. We demonstrate that when the valley degeneracy is broken a charge current is established at the grain boundary interface. These findings provide the working principle of a valley to charge current converter, which is relevant for the emergent field of valleytronics.
\end{abstract}


\maketitle

\section{Introduction}
Dirac materials \cite{diracMat} are condensed matter systems whose low-energy behavior is governed by the Dirac equation. Charge carriers belonging to these materials, of which graphene \cite{graphene} is the prototype, behave like massless quasi-particles propagating at the Fermi velocity instead of the speed of light. This effective behavior can be considered as an emergent property caused by the honeycomb lattice structure containing two atoms inside the unit cell. Dirac materials are not a mere curiosity as they have electrical and mechanical properties of technological interest. For instance, graphene is currently considered as high-mobility transparent electrode for the realization of semiconductor-based photodetectors \cite{applphoto1,applphoto2}. In general, the interest in bidimensional materials originates from their easy integration into industrial production processes. Industrial production of graphene-based devices requires large area graphene synthesis which is nowadays implemented by chemical vapor deposition (CVD) technique. Polycristalline samples are routinely obtained using CVD technique. Polycristalline graphene, differently from exfoliated graphene monolayer, is characterized by the presence of grain boundaries originated by the coalescence of monocrystalline grains \cite{gbnatureexp}. The relative orientation of the crystallographic axes of the coalescing grains depends on the growth process. Grain boundaries, being defective regions, are detrimental for graphene mobility and thus the investigation of the scattering properties of graphene grain boundaries is of utmost relevance. Grain boundary physics in graphene has been the object of intense investigation \cite{gb1,gb2,gb3,gb4,gb5,gb6,gb7}. Scattering theory of graphene grain boundaries has been formulated in Ref. \cite{scgb1,scgb2,scgb3,scgb4,romeogbt}. In particular, in Ref. \cite{romeogbt}, it has been shown that, under appropriate circumstances, a grain boundary can be treated as a linear defect separating two regions with rotated crystallographic axes. The misorientation angle between the different sides of the junction has been taken into account by writing the Dirac equation within a rotated reference frame. According to this model, the transmission properties of a grain boundary can be deduced by using modified boundary conditions for the scattering problem. Indeed, ordinary boundary conditions do not preserve the current density at the interface and thus adequate matching conditions are required. An important consequence of modified boundary conditions is that conductive states with linear dispersion relation and reduced group velocity can nucleate along the graphene/graphene junction. These states are localized in close vicinity of the interface and have been proposed as possible nucleation centers of correlated states in graphene. The existence of zero-energy flat band states with insulating character has also been demonstrated \cite{romeolocst}.\\
Lately it has been theoretically suggested and experimentally verified that Fermi velocity in graphene can be a spatially-varying quantity due to many-body effects, coupling with a substrate and curvature effects \cite{vel0,vel1,vel2,vel3}. All these effects may well occur in graphene grain boundaries. In particular, mechanical stress is generated in the interstitial region between coalescing grains as a consequence of the formation mechanism of a grain boundary. For this reason, the presence of a Fermi velocity gradient is expected at the interface between graphene monocrystalline grains.
Motivated by these arguments, in this work we study the scattering problem at the grain boundary interface assuming a negligible misorientation angle between the crystallographic axes of the two sides of the junction. The scattering problem of a Dirac particle in the presence of a Fermi velocity mismatch has been first formulated in one dimension in Ref. \cite{vel1} and subsequently specialized to describe two-dimensional systems in Ref. \cite{th1,th2,th3}.\\
Here a different technique is used which is a direct generalization of the matching matrix approach developed in Ref. \cite{romeogbt}. Within the mentioned framework, a richer set of boundary conditions for the scattering problem has been derived and the experimental implications of these findings have been carefully discussed. In particular we show that Klein tunneling \cite{klein} suppression, which is a key ingredient in grain boundary physics, is correctly accounted by the proposed theory. We also demonstrate that the Klein tunneling suppression is a necessary condition to observe the conversion of a valley-polarized current into a charge current flowing at the grain boundary interface, the latter conversion being of utmost relevance for valleytronics \cite{valleytronics1,valleytronics2,valleytronics3,valleytronics4,valleytronics5}.\\
The work is thus organized as follows. In Sec. \ref{sec:diracHam} we discuss modifications to the Dirac Hamiltonian induced by the Fermi velocity modulation within the framework of the standard symmetrization technique of quantum mechanics. In Sec. \ref{sec:modelGb} we formulate the Hamiltonian model of a grain boundary junction in the presence of velocity mismatch. We carefully discuss the problem of the current conservation at the interface and the general boundary conditions of the scattering problem. Two families of one-parameter boundary conditions are derived which represent a non-trivial generalization of the boundary conditions usually considered in literature. The scattering problem is discussed in Sec. \ref{sec:scatteringproblem}, where the differential conductance is written in terms of angle-resolved transmittance of the interface. Results are given in Sec. \ref{sec:results}, where analytic expressions for the angle-resolved transmittance are also derived both for type I and II boundary conditions. Differential conductance of the interface is studied as a function of the velocity gradient at the interface. The analysis demonstrates that the Klein tunneling is suppressed by the boundary conditions introduced in this work. Current density distribution in close vicinity of the interface is studied in Sec. \ref{sec:current}, where the conditions to obtain a valley to charge current conversion have been analyzed. Conclusions are given in Sec. \ref{sec:concl}.

\section{Dirac Hamiltonian with space-dependent Fermi velocity}
\label{sec:diracHam}
Low-energy physics of charge carriers in graphene is described by the time-dependent Dirac equation:
\begin{equation}
\label{eq:diracEq}
-i \hbar v \mathbf{\sigma} \cdot \nabla \Psi(\mathbf{r})= i \hbar \partial_{t} \Psi(\mathbf{r}),
\end{equation}
where $v$ represents the Fermi velocity, $\mathbf{\sigma}=(\sigma_x,\sigma_y)$ is the Pauli matrix vector and $\Psi(\mathbf{r})=(\Psi_{A}(\mathbf{r}), \Psi_{B}(\mathbf{r}))^{T}$ is a two-component spinor describing the probability density $|\Psi_{\alpha}(\mathbf{r})|^2$ of finding a quasi-particle inside an infinitesimal volume centered in $\mathbf{r}=(x,y)$ and belonging to the sublattice $\alpha \in \{A,B\}$ of the honeycomb lattice. Differently from a Dirac particle in high energy physics, in condensed matter the propagation velocity $v$ can be a space-dependent quantity and therefore the problem of finding a Hamiltonian model describing this situation arises. According to Ref. \cite{vel1}, the simple substitution $v\rightarrow v(\mathbf{r})$ in Equation (\ref{eq:diracEq}) would lead to a non-Hermitian operator. These theoretical difficulties can be overcome by generalizing the arguments given in Ref. \cite{vel1} to the two-dimensional case as done in Ref. \cite{th1}. Accordingly, we get the following Hamiltonian model
\begin{equation}
\label{eq:diracvel}
-i \hbar \sqrt{v(\mathbf{r})} \mathbf{\sigma} \cdot \nabla [\sqrt{v(\mathbf{r})}\Psi(\mathbf{r})]= i \hbar \partial_{t} \Psi(\mathbf{r}),
\end{equation}
which describes a Dirac particle with a space-dependent propagation velocity $v(\mathbf{r})$. When the Fermi velocity varies only along the $x$-direction, i.e. $v(\mathbf{r})=v(x)$, the translational invariance of the problem along the $y$-direction is preserved. Under this assumption, which is appropriate to describe the grain boundary junction formed at the interface between two micrometric graphene grains, Equation (\ref{eq:diracvel}) can be written in the form:
\begin{equation}
\label{eq:diracvelbis}
-i \hbar v(x) \sigma \cdot \nabla \Psi(\mathbf{r})-\frac{i \hbar}{2} \sigma_x \Big(\partial_x v(x)\Big) \Psi(\mathbf{r})=i \hbar \partial_{t} \Psi(\mathbf{r}),
\end{equation}
which clearly collapses into Equation (\ref{eq:diracEq}) when a uniform Fermi velocity (i.e., $\partial_x v(x)=0$) is considered. The first term of the l.h.s of Equation (\ref{eq:diracvelbis}) corresponds to the simple substitution $v \rightarrow v(x)$ in the l.h.s. of Equation (\ref{eq:diracEq}) and, if taken alone, would give a non-Hermitian model. The second term in the l.h.s. of the same equation provides a cure for the non-Hermitian character of the first term.\\ Interestingly, Equation (\ref{eq:diracvelbis}) can be obtained by simply using the standard symmetrization procedure of quantum mechanics according to which the Dirac equation with Fermi velocity gradient can be written as:
\begin{equation}
\label{eq:symmetrizedH}
-\frac{i \hbar}{2} \{ v(x),\sigma \cdot \nabla \} \Psi(\mathbf{r})=i \hbar \partial_{t} \Psi(\mathbf{r}),
\end{equation}
where $\{A,B\}=AB+BA$ represents the anticommutator between the generic operators $A$ and $B$. Once the charge density is introduced in the form $\rho(\mathbf{r})=|\Psi_{A}(\mathbf{r})|^2+|\Psi_{B}(\mathbf{r})|^2$, the continuity equation takes the usual form
\begin{equation}
\label{eq:continuity}
\partial_{t} \rho(\mathbf{r})+ \overrightarrow{\nabla} \cdot \overrightarrow{\mathcal{J}}(\mathbf{r})=0,
\end{equation}
in which, however, the charge current density components, namely
\begin{eqnarray}
\label{eq:current}
\mathcal{J}_{x}(\mathbf{r})&=& \Psi^{\dag}(\mathbf{r}) v(x)\sigma_{x} \Psi(\mathbf{r})\nonumber\\
\mathcal{J}_{y}(\mathbf{r})&=& \Psi^{\dag}(\mathbf{r}) v(x)\sigma_{y} \Psi(\mathbf{r}),
\end{eqnarray}
are directly affected by space variation of the Fermi velocity. An important implication of these observations is that the first quantization operator
$\widehat{\mathcal{J}}_{x/y}(\mathbf{r})=v(x)\sigma_{x/y}$ is not globally defined along the system.

\section{Model of a grain boundary junction with velocity gradient}
\label{sec:modelGb}
Up to now we have demonstrated that quantum mechanical symmetrization procedure provides a consistent Hamiltonian model which describes a Dirac particle
propagating with a space-dependent Fermi velocity.\\
Hereafter we formulate a simple model of grain boundary which is here schematized as a linear defect located at $x=0$, while translational invariance is assumed along the $y$-direction. The latter assumption is appropriate to describe the conduction properties of grain boundary junctions formed by large grains (i.e., tens of micrometers). Under these assumptions, the Dirac equation $\mathcal{H}\Psi(\mathbf{r})=i \hbar \partial_t \Psi(\mathbf{r})$ can be written in terms of the Hamiltonian operator:
\begin{equation}
\label{eq:diracGb}
\mathcal{H}=-\frac{i \hbar}{2} \{ v(x),\sigma \cdot \nabla \}+U(x),
\end{equation}
where we have introduced the grain boundary potential $U(x)=\mathcal{U}_{gb}\delta(x)+\mathcal{U}_s\mathbb{I}\theta(x)$ which is an Hermitian operator, acting on the sublattice degree of freedom, written in terms of the Dirac delta function $\delta(x)$ and the Heaviside step function $\theta(x)$. Distinct physical meaning has to be attributed to the two terms contributing to $U(x)$. First term in $U(x)$, namely $\mathcal{U}_{gb}\delta(x)$ with $\mathcal{U}_{gb}$ a $2 \times 2$ Hermitian operator, mimics the interface scattering potential at the grain boundary. The second term in $U(x)$, i.e. $\mathcal{U}_s\mathbb{I}\theta(x)$ with $\mathcal{U}_s \in \mathbb{R}$ a c-number, is proportional to the $2 \times 2$ identity operator $\mathbb{I}$ and takes into account band misalignment effects originated by charge transfers at the interface. For the sake of simplicity, a step-like velocity profile $v(x)= \theta(-x)v_L + \theta(x)v_R$ is considered and intervalley scattering processes are neglected within the present approach.\\
Charge conduction along the $x$-direction requires the conservation of the current density at the interface, i.e. $\mathcal{J}_{x}(x=0^{-},y)=\mathcal{J}_{x}(x=0^{+},y)$ with the notation $0^{\pm}$ standing for a positive or negative infinitesimal quantity. Adopting the transfer matrix method, it is possible to relate the wavefunction just after ($x=0^{+}$) and before ($x=0^{-}$) the grain boundary line $x=0$. Accordingly,
wavefunction at the interface obeys the relation
\begin{equation}
\label{eq:matc}
\Psi(0^{+},y)=\mathcal{M}\Psi(0^{-},y)
\end{equation}
with $\mathcal{M}$ a matching matrix which depends on the grain boundary potential $\mathcal{U}_{gb}\delta(x)$.\\
We are interested in determining the structure of admissible matching matrices $\mathcal{M}$ in the absence of a precise knowledge of the grain boundary potential, which is here assumed to be a phenomenological term related to the microscopic structure of the grain boundary. Current density conservation and the wavefunction matching condition provide a constraint for the admissible matching matrices which is written in the form of a matrix equation:
\begin{equation}
\label{eq:matchingeq}
\mathcal{M}^{\dag}\sigma_{x}\mathcal{M}=\gamma \sigma_{x},
\end{equation}
with $\gamma=v_L/v_R$. Equation (\ref{eq:matchingeq}) remains unaffected by the transformation $\mathcal{M} \mapsto e^{i \varphi}\mathcal{M}$ and therefore this property will be used to omit global phase factors in the following discussion. Evident solutions of Equation (\ref{eq:matchingeq}) are $\mathcal{M}=\sqrt{\gamma}\sigma_{x}$ (in virtue of the properties $\sigma_{x}^2=\mathbb{I}$ and $\sigma_{x}=\sigma^{\dag}_{x}$) and $\mathcal{M}=\sqrt{\gamma}\mathbb{I}$. Interestingly, the diagonal matching matrix $\mathcal{M}=\sqrt{\gamma}\mathbb{I}$ is associated with the wavefunction matching $\Psi(0^{+},y)=\sqrt{\gamma}\Psi(0^{-},y)$, which has been extensively used in previous works \cite{vel1,th1,th2}.\\
Finding a general solution of Equation (\ref{eq:matchingeq}) is the object of the following discussion. Let us seek for a solution of Equation (\ref{eq:matchingeq}) in form of a real element matrix $\mathcal{M}$. Matrix equation is equivalent to the following conditions on the matching matrix elements:
\begin{eqnarray}
\label{eq:matchingcomponent}
\mathcal{M}_{11}\mathcal{M}_{21}=0\nonumber\\
\mathcal{M}_{22}\mathcal{M}_{12}=0\nonumber\\
\mathcal{M}_{11}\mathcal{M}_{22}+\mathcal{M}_{12}\mathcal{M}_{21}=\gamma.
\end{eqnarray}
Equations (\ref{eq:matchingcomponent}) define an underdetermined system (three equations and four unknown elements $\mathcal{M}_{ij}$) and thus solutions are parametrized by one phenomenological interface parameter. Solutions can be classified into two types, namely, diagonal matching matrices (type $I$)
\begin{equation}
\mathcal{M}^{(I)}=\sqrt{\gamma}\left(
                    \begin{array}{cc}
                      g(\gamma) & 0 \\
                      0 & g(\gamma)^{-1} \\
                    \end{array}
                  \right)
\end{equation}
and off-diagonal matching matrices (type $II$)
\begin{equation}
\label{eq:mat2}
\mathcal{M}^{(II)}=\sqrt{\gamma}\left(
                    \begin{array}{cc}
                      0 & g(\gamma) \\
                      g(\gamma)^{-1} & 0 \\
                    \end{array}
                  \right),
\end{equation}
with $g(\gamma)=\lambda/\sqrt{\gamma}$ and $\lambda$ an interface parameter such that $\lim_{\mathcal{U}_{gb} \rightarrow 0 } \ \lambda=1$. Once the matching matrices are known, the physical properties of the interface can be studied by solving a scattering problem which will be discussed in the following section.

\section{Scattering problem of a grain boundary junction}
\label{sec:scatteringproblem}

\begin{figure}[h]
	\centering
	\includegraphics[scale=0.4]{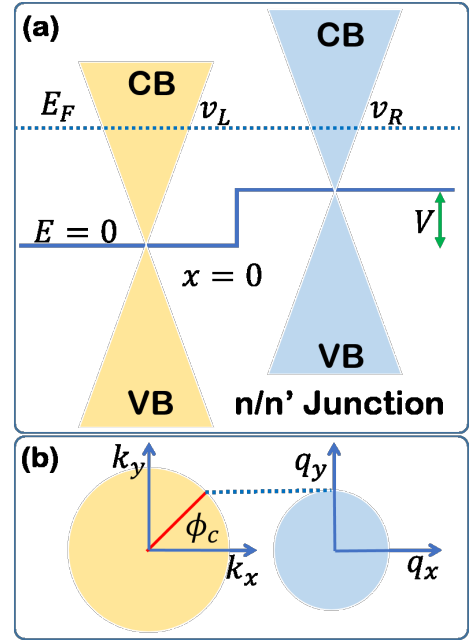}
	\caption{(a) Band alinement of an n/n' graphene junction with velocity mismatch. Conduction and valence bands are labeled by CB and VB, respectively. (b) Critical angle $\phi_c$ above which the particle momentum along the interface cannot be conserved.}
	\label{fig1}
\end{figure}

Let us consider an n/n' grain boundary junction with band alinement depicted in Fig. (\ref{fig1}). An electron coming from the left side of the interface ($x<0$) is described by the scattering wavefunction
\begin{equation}
\label{eq:leftwf}
\Psi_{L}(x,y)=\frac{e^{i k_{y}y}}{\sqrt{2 v^{L}_{x}}}\Bigl\{ \left[
                                                               \begin{array}{c}
                                                                 1 \\
                                                                 e^{i \phi} \\
                                                               \end{array}
                                                             \right] e^{i k_{x}x}+\mathcal{R} \left[
                                                               \begin{array}{c}
                                                                 1 \\
                                                                 -e^{-i \phi} \\
                                                               \end{array}
                                                             \right] e^{-i k_{x}x}
\Bigl\},
\end{equation}
in which $\mathcal{R}$ represents the reflection coefficient, $v^{L}_{x}=v_{L} \cos(\phi)$ is the group velocity along the $x$-direction and $\phi \in (-\pi/2,\pi/2)$ represents the incidence angle of the scattering process. The electron can be transmitted to the right side of the junction ($x>0$) with probability $|\mathcal{T}|^2$, the latter process being described by the scattering wavefunction:
\begin{equation}
\label{eq:rightwf}
\Psi_{R}(x,y)=\mathcal{T}\frac{e^{i q_{y}y}}{\sqrt{2 v^{R}_{x}}}\left[
                                                               \begin{array}{c}
                                                                 1 \\
                                                                 e^{i \phi_t} \\
                                                               \end{array}
                                                             \right] e^{i q_{x}x},
\end{equation}
where we have introduced the $x$-component of the group velocity $v^{R}_{x}=v_{R} \cos(\phi_t)$ and the transmission angle $\phi_t$. The transmission and reflection coefficients, namely $\mathcal{T}$ and $\mathcal{R}$, are not independent due to the particle flux conservation implying the conservation relation $|\mathcal{T}|^2+|\mathcal{R}|^2=1$. The energy dispersion relations on the left and on the right side of the interface are given by
\begin{eqnarray}
E_{L}(\vec{k}) &=& \hbar v_{L}|\vec{k}|\nonumber\\
E_{R}(\vec{q}) &=& \hbar v_{R}|\vec{q}|+V,
\end{eqnarray}
with $V\equiv \mathcal{U}_{s}>0$ the amplitude of the potential step at the interface. Elastic scattering events preserve the energy $E$ and this requires that the equality $E_{L}(\vec{k})=E_{R}(\vec{q})=E$ is preserved. Under the assumption that $E-V>0$, which is consistent with an n/n' junction, one easily obtain $|\vec{k}|=E/(\hbar v_{L})$ and $|\vec{q}|=(E-V)/(\hbar v_{R})$, while the wavevectors are given by $\vec{k}=|\vec{k}|(\cos(\phi),\sin(\phi))$ and $\vec{q}=|\vec{q}|(\cos(\phi_{t}),\sin(\phi_{t}))$. Translational invariance along the $y$-direction implies that the corresponding component of the particle momentum is a conserved quantum number during the scattering event, i.e. $k_{y}=q_{y}$. The conservation of the particle momentum along the interface immediately implies a relation between the incidence and the transmission angle which can be presented in the form
\begin{equation}
\label{eq:transmissionangle}
\phi_{t}=\arcsin \Bigl[\frac{E\sin{\phi}}{\gamma (E-V)}\Bigl].
\end{equation}
The transmission angle $\phi_{t}$ depends on the velocity gradient at the interface and on the energy $E$ of the incident particles, the latter dependence being absent when $V=0$. Particles transmission is possible if the incidence angle fulfills the requirement $|\phi|\leq \phi_{c}$, with
\begin{equation}
\phi_{c}=\arcsin \Bigl [\frac{\gamma (E-V)}{E} \Bigl]
\end{equation}
the critical angle above which the total reflection phenomenon takes place. For highly n-doped graphene/graphene junctions, the relevant scattering energies are typically much greater than the potential step at the interface (i.e., $E\gg V$) and thus the critical angle takes the approximate form $\phi_{c} \approx \arcsin (\gamma)$, which is only affected by the velocity gradient at the interface.\\
Using the scattering wavefunctions (Equations (\ref{eq:leftwf}) and (\ref{eq:rightwf})) and the matching conditions (Equation (\ref{eq:matc})), the angle-resolved scattering coefficients $\mathcal{R}(E, \phi)$ and $\mathcal{T}(E, \phi)$ are obtained. The transmission probability $|\mathcal{T}(E, \phi)|^2$, which is related to the differential conductance of the junction, is strongly affected by the specific matching matrix used to solve the scattering problem. Different structures of the matching matrix correspond to different microscopic properties of the grain boundary interface. Once the scattering problem has  been solved, the zero-temperature differential conductance of the junction can be evaluated according to the following relation:
\begin{equation}
\label{eq:diffcond}
\mathcal{G}=g_{s}g_{v}\frac{e^2}{h}\Big (\frac{k^{L}_{F}W}{2 \pi}\Big ) \int_{-\phi_{c}}^{\phi_{c}} d \phi \Big [|\mathcal{T}(E_F, \phi)|^2 \cos(\phi)\Big],
\end{equation}
where $g_{s}g_{v}=4$ takes into account the spin and the valley degeneracy, $W$ is the transverse dimension of the junction and $k^{L}_{F}=E_{F}/(\hbar v_{L})$ represents the modulus of the Fermi wavevector. For a transparent interface (i.e., $|\mathcal{T}(E_F, \phi)|^2 =1$), which is obtained in the absence of velocity mismatch ($\gamma=1$) and taking $V=0$ and $\mathcal{U}_{gb}=0$, the maximal conductance of the junction, i.e.
\begin{equation}
\mathcal{G}_{max}=g_{s}g_{v}\frac{e^2}{h}\Big (\frac{k^{L}_{F}W}{\pi}\Big ),
\end{equation}
can be written in terms of the available electronic modes $\mathcal{N}(E_F)=g_{s}g_{v}(k^{L}_{F}W) / \pi$ contributing to the charge transport at the Fermi level. In the experiments, the number of electronic modes contributing to the transport can be altered by changing the Fermi level with a back-gate.
Using the property $|\mathcal{T}(E_F, \phi)|^2=|\mathcal{T}(E_F, -\phi)|^2$ in Equation (\ref{eq:diffcond}), one can present the differential conductance in the simple form:
\begin{equation}
\label{eq:diffcondsempl}
\mathcal{G}=\mathcal{G}_{max} \int_{0}^{\phi_{c}} |\mathcal{T}(E_F, \phi)|^2 \cos(\phi) d \phi,
\end{equation}
which contains an additional dependence on $E_F$ hidden in $\mathcal{G}_{max}$ and $\phi_{c}$. Sometimes the transmission probability of an interface presents a negligible dependence on the incidence angle $\phi$ and therefore it is well approximated by $|\mathcal{T}(E_F,\phi)|^2 \approx \Xi \theta(\phi_{c}-|\phi|)$.
Under this assumption, one easily gets:
\begin{equation}
\label{eq:condapprox}
\mathcal{G}/\mathcal{G}_{max} \approx \Xi \sin(\phi_{c})=\Xi \Bigl (1-\frac{V}{E_{F}} \Bigl) \gamma,
\end{equation}
where the quantity $\Xi \in [0,1]$ represents the transmission probability.
As a final comment, we do observe that Equation (\ref{eq:diffcondsempl}) provides a direct link between the scattering properties of the interface and the differential conductance of the system, which is an experimentally accessible quantity.

\section{Results and discussions}
\label{sec:results}
Hereafter, we present the solution of the scattering problem using type I or type II boundary conditions, respectively. We carefully analyze the angle-resolved transmittance of the interface as a function of all relevant parameters. Once the scattering properties of the interface have been characterized, experimental implications of these findings are discussed.

\begin{figure*}[!t]
	\includegraphics[scale=0.9]{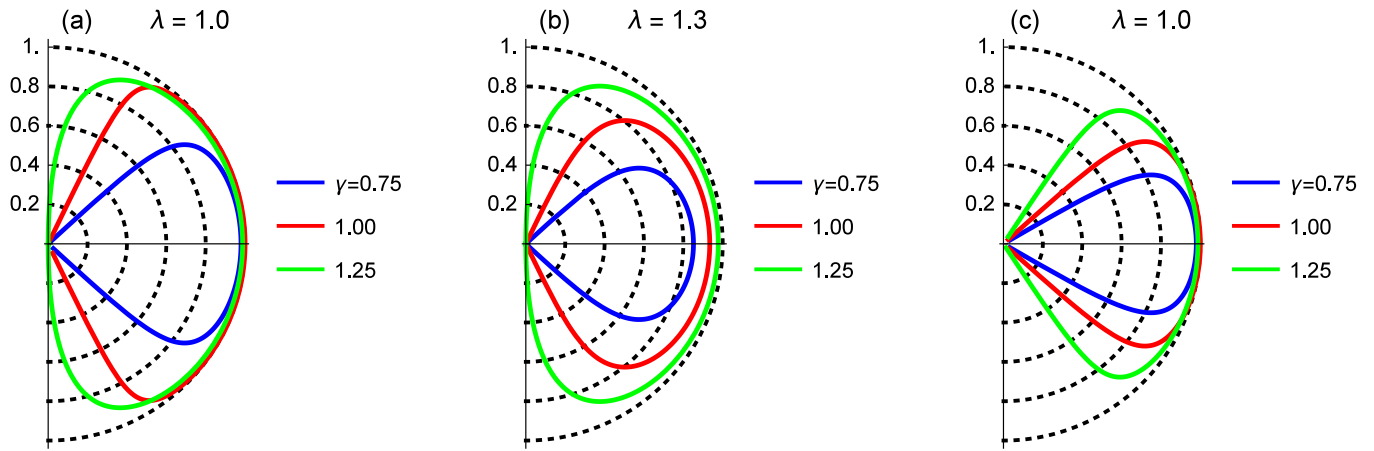}
	\caption{Polar plot of the transmittance $|\mathcal{T}(E_F, \phi)|^2$ as a function of the incidence angle $\phi$ obtained according to Equation (\ref{eq:trtypeone}). Different curves in each panel are obtained by setting different $\gamma$ values. The model parameters are: (a) $\lambda=1.0$ and $V/E_{F}=0.1$; (b) $\lambda=1.3$ and $V/E_{F}=0.1$; (c) $\lambda=1.0$ and $V/E_{F}=0.33$. Normal incidence transmittance is suppressed for $\lambda \neq \gamma^2$, while the critical angle can be altered by changing the potential step amplitude $V$ or the velocity gradient at the interface.}
	\label{fig2}
\end{figure*}

\subsection{Solution of the scattering problem using type I boundary conditions}
We have solved the scattering problem of the n/n' junction depicted in Fig. (\ref{fig1}) by using type I boundary conditions implemented by the matching matrix $\mathcal{M}^{(I)}$. Solution for the angle-resolved transmittance $|\mathcal{T}(E_F, \phi)|^2$ can be presented in the following form:
\begin{equation}
\label{eq:trtypeone}
|\mathcal{T}(E_F, \phi)|^2=\frac{4 \gamma \lambda^2 \cos(\phi)\cos(\phi_{t})}{\gamma^2+\lambda^4+2 \gamma \lambda^{2} \cos(\phi+\phi_{t})},
\end{equation}
where, according to Equation (\ref{eq:transmissionangle}), the transmission angle $\phi_{t}$ depends on the Fermi energy. The analysis of Equation (\ref{eq:trtypeone}) shows that the normal incidence transmittance $|\mathcal{T}(E_F, \phi=0)|^2$ takes the simple energy-independent form
\begin{equation}
\label{eq:zerotrans1}
|\mathcal{T}(E_F, \phi=0)|^2=\frac{4 \lambda^2 \gamma}{(\gamma+\lambda^2)^2},
\end{equation}
which is maximized if $\gamma=\lambda^2$. Interestingly, assuming that $\gamma=\lambda^2$ implies that the matching matrix takes the form $\mathcal{M}^{(I)}=\sqrt{\gamma}\mathbb{I}$, which implements the boundary conditions usually considered in literature. In general $\gamma \neq \lambda^2$ and thus the normal incidence transmittance is reduced compared to its maximal value. Therefore, generalized boundary conditions described in this work correctly reproduce the Klein tunneling suppression, which is an expected feature of the grain boundary transmittance.\\
\begin{figure}[h]
\centering
	\includegraphics[scale=1.0]{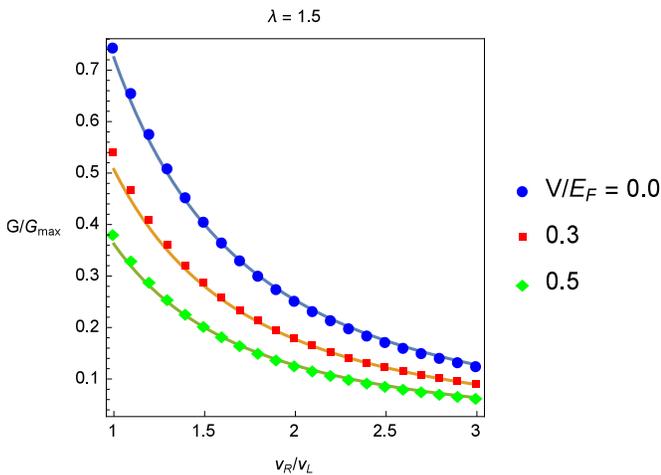}
	\caption{Normalized conductance $\mathcal{G}/\mathcal{G}_{max}$ as a function of $v_R/v_L$ obtained by using Equation (\ref{eq:diffcondsempl}) and (\ref{eq:trtypeone}). Different curves correspond to different potential step values ($V/E_{F}=0.0$ (Circle), $V/E_{F}=0.3$ (Square), $V/E_{F}=0.5$ (Diamond)). Full lines correspond to the approximated expression given in Equation (\ref{eq:fit1cond}) with $\mathcal{A}=0.85$. The interface parameter has been fixed to $\lambda=1.5$.}
	\label{fig3}
\end{figure}
In Figure (\ref{fig2}) we show the $|\mathcal{T}(E_F, \phi)|^2$ \textit{versus} $\phi$ curves deduced by using Equation (\ref{eq:trtypeone}) for different values of the model parameters. The analysis of the different panels evidences that the normal incidence transmittance is suppressed for $\gamma \neq \lambda^2$, while the critical angle $\phi_c$ can be altered by changing the potential step amplitude $V$ or the velocity gradient at the interface, the latter being controlled by $\gamma$. As a general comment we observe that the angle-resolved transmittance is almost constant as a function of the incidence angle $|\phi|<\phi_{c}$, while a substantial variation is observed in close vicinity of the critical angle $\pm \phi_{c}$. These features are compatible with the assumptions under which we have derived Equation (\ref{eq:condapprox}) and thus we expect that the conductance curves, obtained by using Equation (\ref{eq:diffcondsempl}), follow the approximated behavior given by
\begin{figure*}[!t]
	\includegraphics[scale=0.9]{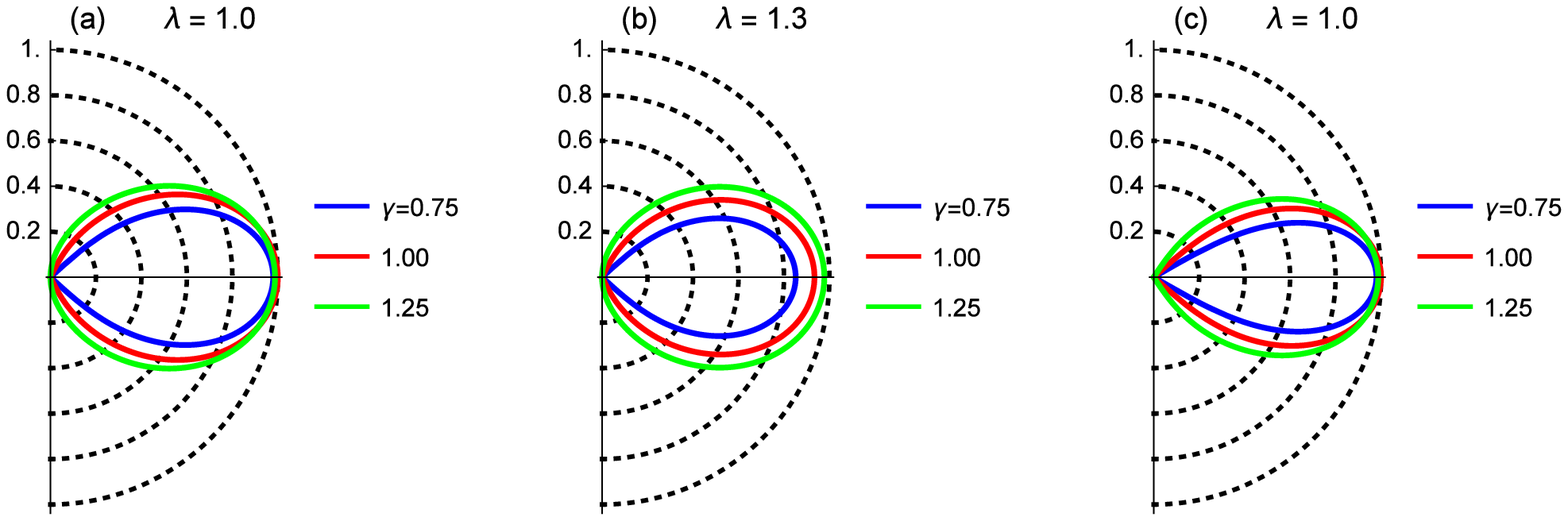}
	\caption{Polar plot of the transmittance $|\mathcal{T}(E_F, \phi)|^2$ as a function of the incidence angle $\phi$ obtained according to Equation (\ref{eq:trtypetwo}). Different curves in each panel are obtained by setting different $\gamma$ values. The model parameters are: (a) $\lambda=1.0$ and $V/E_{F}=0.1$; (b) $\lambda=1.3$ and $V/E_{F}=0.1$; (c) $\lambda=1.0$ and $V/E_{F}=0.33$.}
	\label{fig4}
\end{figure*}
\begin{equation}
\label{eq:fit1cond}
\mathcal{G}/\mathcal{G}_{max} = \mathcal{A} \ \Bigl (\frac{2 \lambda \gamma}{\gamma+\lambda^2}\Bigl )^2\Bigl (1-\frac{V}{E_{F}} \Bigl),
\end{equation}
where a fitting parameter $0<\mathcal{A}<1$ has been introduced. In order to verify the above conclusions, in Figure (\ref{fig3}) we show the conductance curves as a function of the velocity ratio $v_R/v_L$ for different values of the step potential amplitude $V$. Continuous curves are obtained by using Equation (\ref{eq:fit1cond}), while symbols are used to evidence conductance values computed by using Equation (\ref{eq:diffcondsempl}). By direct inspection of Figure (\ref{fig3}), the agreement between the approximated expression and the exact conductance values is evident. Conductance curves are strongly affected by the velocity gradient at the interface and show a decreasing behavior with respect to $v_R/v_L$ which is described by the functional form
\begin{equation}
\label{eq:fitcond2}
\mathcal{G}/\mathcal{G}_{max} \sim \frac{4 \lambda^2}{(1+\frac{v_R}{v_L}\lambda^2)^2}.
\end{equation}
The conductance lowering described by Equation (\ref{eq:fitcond2}) is much faster than the one induced by the $\mathcal{G}/\mathcal{G}_{max} \sim v_L/v_R$  behavior which is obtained by using standard boundary conditions implemented by $\mathcal{M}^{(I)}=\sqrt{\gamma}\mathbb{I}$.

\subsection{Solution of the scattering problem using type II boundary conditions}

We have solved the scattering problem of the n/n' junction using type II boundary conditions implemented via the matching matrix $\mathcal{M}^{(II)}$ reported in Equation (\ref{eq:mat2}). Once the scattering problem has been solved, the angle-resolved transmittance $|\mathcal{T}(E_F, \phi)|^2$ can be presented in the following form:
\begin{equation}
\label{eq:trtypetwo}
|\mathcal{T}(E_F, \phi)|^2=\frac{4 \gamma \lambda^2 \cos(\phi)\cos(\phi_{t})}{\gamma^2+\lambda^4+2 \gamma \lambda^{2} \cos(\phi-\phi_{t})},
\end{equation}
which differs from Equation (\ref{eq:trtypeone}) just for a sign inside the cosine argument at the denominator. Despite this difference, the normal incidence transmittance $|\mathcal{T}(E_F, \phi=0)|^2$ maintains the same analytic form presented in Equation (\ref{eq:zerotrans1}) and consequently the same properties. The general aspect of the transmittance curves is presented in Figure (\ref{fig4}), where the model parameters have been fixed as done in Figure (\ref{fig2}). The analysis of the transmittance curves shows that type II boundary conditions describe more opaque interfaces compared to the type I case. Indeed, the transmission probability of scattering events with high incidence angle is strongly suppressed, the latter phenomenon being only weakly affected by the velocity gradient at the interface.\\
\begin{figure}[h]
\centering
	\includegraphics[scale=1.0]{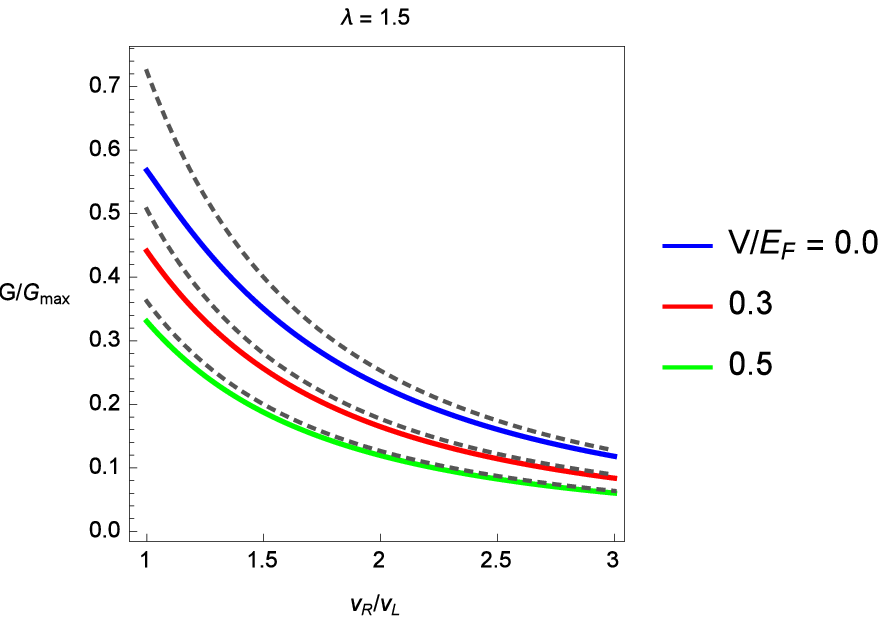}
	\caption{Normalized conductance $\mathcal{G}/\mathcal{G}_{max}$ as a function of $v_R/v_L$ obtained by using Equation (\ref{eq:diffcondsempl}) and (\ref{eq:trtypetwo}). Different curves correspond to different potential step values ($V/E_{F}=0.0$ (blue full line), $V/E_{F}=0.3$ (red full line), $V/E_{F}=0.5$ (green full line)). Dashed lines correspond to conductance curves presented in Figure (\ref{fig3}) and obtained by using the type I boundary conditions. The interface parameter has been fixed to $\lambda=1.5$.}
	\label{fig5}
\end{figure}
Conductance properties of the interface as a function of $v_R/v_L$ are studied in Figure (\ref{fig5}) with the same parameters choice of Figure (\ref{fig3}). Dashed curves represent the conductance curves calculated by using type I boundary conditions and reported in Figure (\ref{fig3}), while full lines are used for type II conductance curves. The comparison between type I and type II conductance curves, namely $(\mathcal{G}/\mathcal{G}_{max})_{I}$ and $(\mathcal{G}/\mathcal{G}_{max})_{II}$, shows that $(\mathcal{G}/\mathcal{G}_{max})_{I} \geq (\mathcal{G}/\mathcal{G}_{max})_{II}$. Type I and type II conductance curves tend to become indistinguishable when the critical angle $\phi_{c}$ goes to zero, the latter condition being satisfied when $v_R/v_L \gtrsim 2$ or $E_F \approx V$. Under the mentioned assumptions, type I and type II conductance curves collapse on each other and they are well approximated by Equation (\ref{eq:fit1cond}).

\section{Analysis of the current density at the interface}
\label{sec:current}
A further characterization of the interface properties can be obtained by studying the current density distribution at the interface. It is convenient to work with energy eigenfunctions $\psi_{E}(x,y)$, which correspond to stationary solutions $\Psi_{E}(x,y,t)=\psi_{E}(x,y) e^{-i Et/\hbar}$ of the Dirac equation. Due to the presence of the interface in $x=0$, the energy eigenfunctions can be decomposed as $\psi_{E}(x,y)=\psi_{L}(x,y)\theta(-x)+\psi_{R}(x,y)\theta(x)$ with
\begin{eqnarray}
\label{eq:energyeigen}
\psi_{L}(x,y)&=&\frac{e^{i k_{y}y}}{\sqrt{2}}\Bigl\{ \left[
                                                               \begin{array}{c}
                                                                 1 \\
                                                                 e^{i \phi} \\
                                                               \end{array}
                                                             \right] e^{i k_{x}x}+\mathcal{A} \left[
                                                               \begin{array}{c}
                                                                 1 \\
                                                                 -e^{-i \phi} \\
                                                               \end{array}
                                                             \right] e^{-i k_{x}x}
\Bigl\}\nonumber\\
\psi_{R}(x,y)&=&\mathcal{B}\frac{e^{i k_{y}y}}{\sqrt{2}}\left[
                                                               \begin{array}{c}
                                                                 1 \\
                                                                 e^{i \phi_t} \\
                                                               \end{array}
                                                             \right] e^{i q_{x}x},
\end{eqnarray}
where the momentum-dependent coefficients $\mathcal{A}$ and $\mathcal{B}$, which are related to $\mathcal{R}$ and $\mathcal{T}$ determined before, are fixed by the boundary conditions. In writing Equation (\ref{eq:energyeigen}), a quantum state with positive group velocity along the $x$-direction has been considered. The latter describes matter waves incident on the grain boundary from the left side. Using $\psi_{E}(x,y)$ the expectation values of the current density components, namely $J_{x/y}=\psi_{E}(x,y)^{\dag}v(x)\sigma_{x/y}\psi_{E}(x,y)$, can be obtained. After direct computation, we get the $x$-component of the current density, namely
\begin{eqnarray}
J_{x}(x,y)=\theta(-x) \Big[1-|\mathcal{A}|^2 \Big] v_{L}\cos(\phi)+\theta(x) |\mathcal{B}|^2 v_{R} \cos(\phi_t), \nonumber
\end{eqnarray}
and the $y$-component
\begin{eqnarray}
J_{y}(x,y)&=&\theta(-x) \Big [(1+|\mathcal{A}|^2)v_{L}\sin(\phi)+\Upsilon_{\phi}(x)\Big]+\nonumber\\
&+&\theta(x) |\mathcal{B}|^2 v_{R} \sin(\phi_t)\nonumber,
\end{eqnarray}
where we have introduced the notation
\begin{equation}
\Upsilon_{\phi}(x)=2 v_{L}\mathrm{Im}[\mathcal{A}^{\ast}e^{i(\phi+2k_{x}x)}].
\end{equation}
\begin{figure}[h]
\centering
\vspace{0.5 cm}
	\includegraphics[scale=1.5]{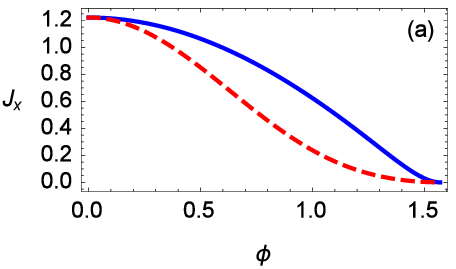}\\
    \includegraphics[scale=1.5]{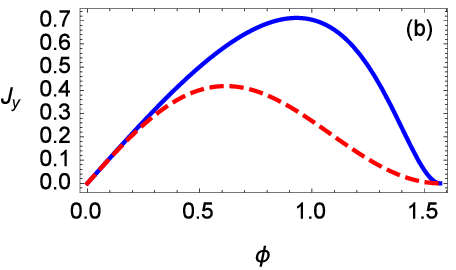}
	\caption{Panel (a): $x$-component of the current density, $J_{x}$ (in units of $v_R$), as a function of the incidence angle $\phi$. The property $J_{x}(\phi)=J_{x}(-\phi)$ has been exploited to reduce the visualization range. Panel (b): $y$-component of the current density, $J_{y}$ (in units of $v_R$), as a function of the incidence angle $\phi$. The property $-J_{x}(\phi)=J_{x}(-\phi)$ has been exploited to reduce the visualization range. In each panel, different curves are obtained by using type I (full line) or type II (dashed line) boundary conditions. The interface model parameters have been fixed as: $\lambda=1.3$, $\gamma=1.25$, $V/E_{F}=0.1$.}
	\label{fig6}
\end{figure}
The conservation of the $x$-component of the current density at the interface, i.e. $J_{x}(0^{+},y)=J_{x}(0^{-},y)$, immediately implies the relations
\begin{eqnarray}
|\mathcal{R}|^2 &=& |\mathcal{A}|^2\nonumber\\
|\mathcal{T}|^2 &=& |\mathcal{B}|^2 \Bigl(\frac{v_{R}\cos(\phi_{t})}{v_{L}\cos(\phi)}\Bigl).
\end{eqnarray}
Moreover, the translational invariance of the problem along the $y$-direction implies that the current density $\vec{J}=(J_{x}, J_{y})$ does not depend on this variable, while the stationary condition $\vec{\nabla}\cdot \vec{J}=0$ is respected. The current density just derived represents the elementary contribution provided by a scattering event with fixed energy $E$ and incidence angle $\phi$. Accordingly, the current density depends on these variables. The current density components on the right side of the grain boundary ($x>0$) are analyzed in Figure \ref{fig6}. Current density components are spatially uniform and present a dependence on the incidence angle $\phi$. In particular, the $x$-component of the current density (Figure \ref{fig6} (a)) is an even function of the incidence angle, while $y$-component (Figure \ref{fig6} (b)) presents an odd dependence on $\phi$. Different curves in Figure \ref{fig6} are obtained by using type I (full line) or type II (dashed line) boundary conditions and show that type II boundary conditions induce a pronounced suppression of $J_x$ at high incidence angles (see Figure \ref{fig6} (a)). The average contribution to the $\alpha \in \{x,y\}$ component of the current density due to a scattering process with assigned energy $E$ (close to the Fermi energy) and random incidence angle $\phi \in (-\pi/2,\pi/2)$ can be evaluated as
\begin{equation}
\overline{J_{\alpha}}=\frac{1}{\pi}\int_{-\pi/2}^{\pi/2}J_{\alpha}(x,y) d \phi,
\end{equation}
where an uniform distribution of incidence angles has been assumed. A qualitative analysis shows that $\overline{J_{x}}$ presents a sensitive dependence on the boundary conditions. Such a dependence corroborates our previous conclusion that type II interfaces are less conductive than those described by type I boundary conditions. A similar analysis on $\overline{J_{y}}$ shows that a single term survives the angular averaging giving rise to the following unexpected relation:
\begin{equation}
\label{eq:intcurr}
\overline{J_{y}}=\frac{\theta(-x)}{\pi}\int_{-\pi/2}^{\pi/2} \Upsilon_{\phi}(x) d \phi.
\end{equation}
Interestingly there are no fundamental reasons for which the $y$-component of the current density should be conserved at the interface and, accordingly,
we arrive at the counterintuitive conclusion that, in principle, somewhere $\overline{J_{y}} \neq 0$. Consistently with our previous observations, Equation (\ref{eq:intcurr}) implies that for $x>0$ the $y$-component of the current density does not contribute to the particle flux through the interface (i.e., $\overline{J_{y}}=0$). To help the intuition, we rewrite Equation (\ref{eq:intcurr}) in a more explicit form
\begin{equation}
\label{eq:intcurrsempl}
\overline{J_{y}}=\frac{2 v_{L}\theta(-x)}{\pi}\int_{-\pi/2}^{\pi/2} |\mathcal{A}|\sin \Big[\phi-\varphi_{\mathcal{A}}+\frac{2 E x}{\hbar v_{L}}\cos(\phi)\Big ] d \phi,
\end{equation}
where we have introduced the $\phi$-dependent quantities $|\mathcal{A}|$ and $\varphi_{\mathcal{A}}$ and the notation $\mathcal{A}=|\mathcal{A}|e^{i \varphi_{\mathcal{A}}}$. Observing that $|\mathcal{A}|=|\mathcal{R}|$, Equation (\ref{eq:intcurrsempl}) explicitly shows that, under appropriate conditions, a non-vanishing $\overline{J_{y}}$ is generated by the interference of back-scattered and incoming matter waves. A further progress can be made by assuming that the quantity $\mathcal{A}$ presents a weak dependence on the incidence angle $\phi$. Under this assumption, which is sometimes reasonable, the $y$-component of the current density for $x<0$ takes the form
\begin{equation}
\label{eq:intcurrspecial}
\overline{J_{y}} \approx 2 v_{L}|\mathcal{A}| \Big[J_{1}\Big(\frac{2 E x}{\hbar v_{L}}\Big)\cos(\varphi_{\mathcal{A}})-H_{-1}\Big(\frac{2 E x}{\hbar v_{L}}\Big)\sin(\varphi_{\mathcal{A}}) \Big ],
\end{equation}
the latter being expressed in terms of the Bessel function $J_{1}(z)$ and the Struve function $H_{-1}(z)$. Both the Struve and the Bessel function are characterized by an oscillating behavior going to zero as the argument $z\rightarrow -\infty$. The oscillation wavelength is a fraction of the Fermi wavelength $\lambda_{F}$ which, on its turn, depends on the Fermi energy. From the above arguments, we reach the conclusion that $\overline{J_{y}}$ is localized along the grain boundary interface ($x=0$). The asymptotic behavior of the special functions for $x\rightarrow -\infty$ suggests that $\overline{J_{y}}$ goes to zero as $|x|^{-1/2}$. At finite temperature, however, Equation (\ref{eq:intcurrspecial}) have to be averaged over an energy window centered at the Fermi energy $E_F$ with amplitude $2\Delta \sim k_{B}T$ determined by the thermal energy. This averaging procedure is required to take into account all scattering processes contributing to $\overline{J_{y}}$. Accordingly, one can estimate that at finite temperature the $y$-component of the current density goes to zero as $|x|^{-3/2}$. The latter result can be easily proven by using the properties of the special functions under integration. From the above arguments, we conclude that $\overline{J_{y}}$, if present, flows along the grain boundary interface and penetrates the bulk for a distance comparable to $\lambda_{F}$.\\
So far we have provided a qualitative description of the expected physical properties of $\overline{J_{y}}$. Hereafter we refine the analysis to discuss the effect of the different boundary conditions and the conditions under which a non-vanishing interface current $\overline{J_{y}}$ exits. First of all, we rewrite $\overline{J_{y}}$ as
\begin{equation}
\label{eq:intcurrnum}
\overline{J_{y}}=\frac{\theta(-x)}{\pi}\int_{0}^{\pi/2} \mathcal{C}_{\phi}(x) d \phi,
\end{equation}
where we have introduced the auxiliary function $\mathcal{C}_{\phi}(x)=\Upsilon_{\phi}(x)+\Upsilon_{-\phi}(x)$ which is the central object of our subsequent discussion. Direct computation shows that the auxiliary function takes the following form
\begin{equation}
\label{eq:auxfunc}
\mathcal{C}_{\phi}=\sigma \frac{4 v_{L}(\gamma^2- \lambda^4) \cos(\phi)\sin(z \cos(\phi))}{\gamma^2+\lambda^4+2 \gamma \lambda^{2} \cos(\phi+\sigma \phi_{t})},
\end{equation}
with $\sigma=+1$ ($\sigma=-1$) for type I (type II) boundary conditions and $z=2 E x/(\hbar v_{L})$. Equation (\ref{eq:auxfunc}) implies that a non-vanishing current $\overline{J_{y}}$ exists only when $\gamma \neq \lambda^2$. Thus, a necessary requirement to have $\overline{J_{y}} \neq 0$ is the suppression of the normal-incidence transmittance at the interface, the latter being a fingerprint of the generalized boundary conditions introduced in this work. From the physical viewpoint, it is expected that a grain boundary junction might satisfy the conditions to observe non-vanishing values of $\overline{J_{y}}$. Interestingly, the current direction at the interface is decided by the sign of the quantity $(\gamma^2- \lambda^4)$. For an arbitrary parameter choice, the space-distribution of $\overline{J_{y}}$ can be studied by numerical integration of Equation (\ref{eq:intcurrnum}). Some analytical progress can be made when opaque interfaces are considered. The limit of opaque interface can be studied by considering Equation (\ref{eq:intcurrnum}) and (\ref{eq:auxfunc}) under the assumptions $\lambda \rightarrow 0$ or $\lambda \rightarrow \infty$. These limits, which correspond to distinct sublattice matchings, have not to be meant in a strict mathematical sense, rather they indicate a condition under which the interface conductance is strongly suppressed. When the limit $\lambda \rightarrow 0$ is considered, we get:
\begin{equation}
\overline{J_{y}}=2 \sigma v_{L} J_{1}\Big(\frac{2 E x}{\hbar v_{L}}\Big) \theta(-x),
\end{equation}
while a current with opposite sign is obtained when the limit $\lambda \rightarrow \infty$ is considered. The energy average of $\overline{J_{y}}$ over the relevant energy window $(E_F-\Delta, E_F+\Delta)$ takes the following form:
\begin{eqnarray}
\langle J_{y}\rangle_{E}&=&\frac{1}{2 \Delta}\int_{E_{F}-\Delta}^{E_{F}+\Delta} \overline{J_{y}} d E=\\
&=&\frac{ \sigma v_{L} \theta(-x)}{ z_{F}\epsilon} \Big [J_{0}\Big(z_{F}(1-\epsilon)\Big)-J_{0}\Big(z_{F}(1+\epsilon)\Big)\Big],\nonumber
\end{eqnarray}
where we have introduced the shortened notation $z_{F}=(2 E_{F} x)/(\hbar v_{L})$ and $\epsilon=\Delta/E_{F}$. The asymptotic behavior of $\langle J_{y}\rangle_{E}$ at relevant distance from the interface (i.e. $x \rightarrow -\infty$) is easily deduced by using the asymptotic approximation of the zeroth-order Bessel function. Accordingly, $\langle J_{y}\rangle_{E}$ presents a non-exponential decay $ \sim |x|^{-3/2}$ which confirms our preliminary observations.
\begin{figure}[h]
\centering
	\includegraphics[scale=0.95]{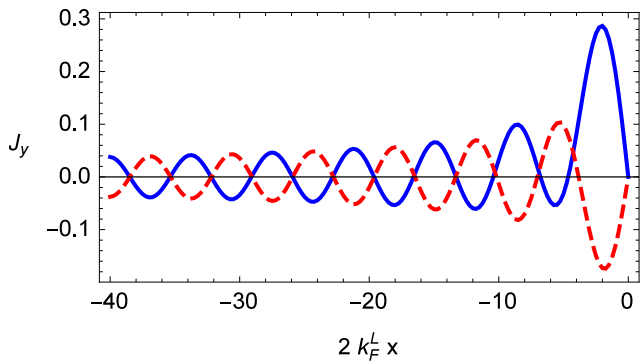}
	\caption{ Interface current $J_{y}$ (in units of $v_{L}$) computed according to Equations (\ref{eq:intcurrnum})-(\ref{eq:auxfunc}) by using type I (full line) or type II (dashed line) boundary conditions. The interface model parameters have been fixed as: $\lambda=1.3$, $\gamma=1.25$, $V/E_{F}=0.1$.}
	\label{fig7}
\end{figure}
The behavior of the interface current for an arbitrary parameters choice can be obtained by numerical integration of Equations (\ref{eq:intcurrnum})-(\ref{eq:auxfunc}). The results of this analysis are shown in Figure (\ref{fig7}) and confirm the conclusions we have obtained by discussing special cases. In particular, depending on the boundary conditions imposed by the grain boundary, both the sign and the magnitude of the interface current are modified. Moreover, as expected, the main contribution to the interface current is localized inside the region defined by the approximate condition $|x|\lesssim \lambda_{F}/\pi$.\\
So far we have discussed the conditions required to observe an interface current while not discussing the physical origin of this current. Indeed, the observation of a non-vanishing interface current requires the existence of a preferential orientation of the particle flux along the interface. This orientation is fixed by the type of boundary condition considered and by the strength of the interface parameter $\lambda$ (to be compared with $\sqrt{\gamma}$). In Ref. \cite{romeolocst} it has been demonstrated that localized defect states with conductive character can nucleate in close vicinity of a linear defect of a Dirac material. Such states, under appropriate circumstances, can exhibit protection from back-scattering events due to the momentum-valley locking. The momentum-valley locking implies that if a conducting defect state with momentum $q$ (parallel to the interface) and valley quantum number $\xi=+1$ exists, it is not accompanied by a state with quantum numbers $-q$ and $\xi=+1$. A defect state with momentum $-q$ is instead present when the valley quantum number $\xi=-1$ is considered. The above scenario explains our findings. Indeed, we have solved a scattering problem by implicitly fixing the valley quantum number $\xi=+1$. Due to the structure of the scattering potential, a conducting defect state with momentum $q$ exists. This state can only support a unidirectional transport along the interface, the latter being responsible for a non-vanishing interface current. Thus the unidirectional defect state is fed by scattering processes involving particles with valley quantum number $\xi=+1$. On the other hand, when scattering processes involving particles with opposite valley quantum number ($\xi=-1$) are considered, we expect to find an interface current with opposite sign compared to the previous case. The latter current is sustained by a defect state with quantum numbers $-q$ and $\xi=-1$. In principle, contributions originated by different valleys can compensate to give a vanishing interface current.\\
In real systems, the current flowing through the grain boundary is originated by a statistical mixture of particles with opposite valley quantum number (i.e. $\xi=+1$ and $\xi=-1$). Homogeneous and defect-free graphene sheets are valley-degenerate systems in which no imbalance is expected between carriers with different valley quantum number. In the presence of grain boundaries, valley degeneracy can be broken and consequently valley-polarized currents can be generated. In particular, it has been theoretically \cite{stegmann2018,carillo2016,settnes2016,milovanovic2016} and experimentally \cite{georgi-exp2017} demonstrated that local strain effects, which are relevant in the presence of grain boundaries, or interaction with a substrate may originate valley-polarized particle currents. Under this condition, we expect that the interface current is not exactly canceled when the valley quantum number is taken into account. For these reasons, we conclude that a non-vanishing charge current at the grain boundary interface is originated by a valley-polarized current flowing through the interface. The above mechanism represents the physical principle of a valley to charge current converter, which is relevant in valleytronics.\\
From the experimental viewpoint, the detection of the interface current described in this work can be performed by means of the current flow imaging technique reported in Ref. \cite{imagingcurr} which is able to provide information about transport phenomena in real space.

\section{Conclusions}
\label{sec:concl}
We have formulated a model of grain boundary junction in graphene in the presence of Fermi velocity mismatch at the interface. The model requires the generalization of the Dirac equation to the case of space-dependent particle propagation velocity which is here implemented by standard symmetrization procedure of quantum mechanical Hamiltonian. We have demonstrated that the Hamiltonian model obtained within the proposed approach is coincident with the one already considered in literature. After the derivation of the continuity equation for the charge current, we have studied general boundary conditions for the scattering problem by adopting the matching matrix method. We have found two different families of boundary conditions parametrized by a single interface parameter. We have discussed the implications of the different boundary conditions on the differential conductance of a grain boundary junction showing the relevant role of the velocity gradient in determining the transmission properties of the interface. Interestingly, the boundary conditions derived in this work provide a suppression of the Klein tunneling, which is an expected feature of a grain boundary interface. Different boundary conditions are accompanied by a peculiar behavior of the angle-resolved transmittance, which is a quantity that directly affects the differential conductance. In particular, we have shown that type I boundary conditions are associated with more conductive interfaces compared to those described by type II boundary conditions. This difference in the conduction properties depends on the fact the type II boundary conditions strongly suppress tunneling processes with high incidence angles that instead  are only limited by the critical angle for type I interfaces. We have analyzed the currents distribution in close vicinity of the grain boundary interface and we have demonstrated the existence of a valley to charge current conversion mechanism, which is of interest for the emergent field of valleytronics. We expect that these findings are relevant for the characterization of the grain boundary physics in Dirac materials.

\section*{Acknowledgment}
Discussions with R. De Luca and M. Salerno are gratefully acknowledged.

\end{document}